\documentstyle[aps,epsfig,prl]{revtex}  
\newcommand{\up}{\uparrow}
\newcommand{\dn}{\downarrow}
\begin{document}    
\twocolumn
[\hsize\textwidth\columnwidth\hsize\csname@twocolumnfalse\endcsname

\title{Effect of the $W$-term for a  $t-U-W$ Hubbard ladder}

\author{ S.\ Daul $^{1}$, D. J. Scalapino $^{2}$ and Steven R. White $^3$ }

\address{ 
$^1$ Institute for Theoretical Physics, University of California, \\
                  Santa-Barbara CA 93106. \\
$^2$ Physics Department, University of California, \\
                  Santa-Barbara CA 93106. \\
$^3$ Department of Physics and Astronomy, University of California, \\
	Irvine CA 92697.
}
\maketitle

\begin{abstract}               

Antiferromagnetic and $d_{x^2-y^2}$-pairing correlations appear delicately
balanced in the 2D Hubbard model. 
Whether doping can tip the balance to pairing is unclear and models with
additional interaction terms have been studied.
In one of these, the square of a local hopping kinetic energy $H_W$ was 
found to favor pairing.
However, such a term can be separated into a number of simpler processes
and one would like to know which of these terms are responsible
for enhancing the pairing.
Here we analyze these processes for a 2-leg Hubbard ladder.

\end{abstract}

\vspace*{2mm}
]


The interplay of antiferromagnetism and $d_{x^2-y^2}$ superconductivity
in the 2D Hubbard model remains an open question. \cite{RecentConf}
Weak coupling calculations originally suggested that doping could drive
the ground state from an antiferromagnet to a $d_{x^2-y^2}$ superconductor.
\cite{Scalapino}
However, numerical Monte Carlo calculations have found only short range 
$d_{x^2-y^2}$ pairing correlations. 
\cite{Imada,Moreo,Gubernatis}
This may be due to the finite lattice sizes that have been studied, the 
difficulty in attaining low temperature results or possibly that the
$t-U$ Hubbard model lies just outside the superconducting parameter regime.

One approach to this problem is then to add various terms to the basic Hubbard 
model and see what it takes to drive it into a superconduting state.
In this spirit, a recent Monte Carlo study \cite{Assaad}
 added a term $H_W$, 
involving the square of the local hopping kinetic energy around a site,
\begin{equation}
      H_W =  -W \sum_i K_i^2
\label{eq:HW}
\end{equation}
with $K_i$ equal to the local kinetic energy involving site $i$ and its
near neighbors at $i+\delta$,
\begin{equation}
        K_i = \sum_{\delta, \sigma=\up,\dn} \left( c^\dag_{i,\sigma}
       c_{i+\delta, \sigma} + 
        c^\dag_{i+\delta, \sigma} c_{i, \sigma}   \right) .
\label{eq:kinetic}
\end{equation}
With $H_W$ added to the 2D $t-U$ Hubbard model, the half-filled system
exhibited a transition from an antiferromagnetic phase to a 
$d_{x^2-y^2}$-pairing phase at a critical value of $W$.
Separating $H_W$ into various pieces, it was found that it contained 
one-electron hopping terms, exchange interactions and triplet and singlet
four particle scattering terms.
One would like to understand which of these terms or what combination
of the terms are responsible for enhancing superconductivity.
Unfortunately because of the fermion sign problem it has not been possible
to carry out a Monte Carlo calculations for the individual terms.
However, density matrix renormalization group (DMRG) techniques 
\cite{White92} can be used to 
study the individual pieces of the $H_W$ interaction. 
Here we describe the results of such a study for a 2-leg ladder.
For such a system, we can determine the effect of the individual terms for
both the half-filled and the doped system.
As we will discuss in the conclusion, it is important to note that the
half-filled 2-leg ladder has a spin gap which distinguishes it from the 2D
half-filled Hubbard model.
Nevertheless, it is instructive to see what effect the various
parts of $W$ have on the pairing correlations for a ladder.

We begin with the usual Hubbard Hamiltonian
\begin{equation}
    H_U = -t \sum_{<ij>,\sigma = \up,\dn} 
   \left( c^\dag_{i,\sigma} c_{j, \sigma} + 
        c^\dag_{j, \sigma} c_{i, \sigma}   \right)
      + U \sum_i n_{i\up} n_{i\dn}
\end{equation}
with a one electron hopping kinetic energy and an onsite Coulomb 
interaction $U$.
The sum $<ij>$ is over all pairs of nearest neighbors.
We will measure all energies in units of $t$.
We then add the interaction (\ref{eq:HW}) with $W$ positive. 
Monte Carlo calculations for a 2D half-filled system with the Hamiltonian
\begin{equation}
      H =  H_U + H_W
\label{eq:tot_H}
\end{equation}
find a quantum phase transition between an antiferromagnetic Mott insulator
and a $d_{x^2-y^2}$-wave superconducting phase
when $W$ is increased to a value of order 0.35.  \cite{Assaad} 
However, the 2D Hubbard model at half-filling has an antiferromagnetic 
ground state while a 2-leg ladder is characterized by a spin gap.
\cite{SpinGap}
Thus, as we will see, the behavior of a two-leg ladder as $W$ is turned on, 
can be different.

It is convenient to decompose the interaction $H_W$ as follows \cite{Assaad}
\begin{equation}
      H_W = \sum_i H_{W_i}
\end{equation}
with
\setcounter{equation}{0}
\renewcommand{\theequation}{6\alph{equation}}
\begin{eqnarray}
  &&	H_{W_1} =  -4 W_1 \sum_i (n_{i\up}  + n_{i\dn} )    \\ 
  &&	H_{W_2} =   -W_2 \sum_{i,\delta,\delta'} \sum_\sigma  
	             c^\dag_{i+\delta, \sigma} c_{i+\delta', \sigma} \\        
  &&	H_{W_3} =  -W_3 \sum_{i,\delta,\delta'} \sum_\sigma \left(  
	             c^\dag_{i, \sigma} c^\dag_{i, -\sigma}
	             c_{i+\delta', -\sigma} c_{i+\delta, \sigma}
	          + \mbox{h.c.} \right)    \\ 
  &&	H_{W_4} =  + W_4  \sum_{i,\delta,\delta'} \left(
             T^\dag_{i\delta',1}T_{i\delta,1} +
             T^\dag_{i\delta',-1}T_{i\delta,-1} +
	     T^\dag_{i\delta',0}T_{i\delta,0}  \right) \nonumber \\ 
  &&	H_{W_5} = -W_5 \sum_{i,\delta} \Delta^\dag_{i\delta}\Delta_{i\delta}  \\ 
  &&  	H_{W_6} = -W_6 \sum_{i,\delta\neq\delta'}
		 \Delta^\dag_{i\delta}\Delta_{i\delta'} . 
\end{eqnarray}
\setcounter{equation}{6}
\renewcommand{\theequation}{\arabic{equation}}
\hspace{-5mm}
Here $ T^\dag_{i\delta,1} = c^\dag_{i,\up} c^\dag_{i+\delta,\up} $, 
$ T^\dag_{i\delta,-1} = c^\dag_{i,\dn} c^\dag_{i+\delta,\dn} $, 
 $ T^\dag_{i\delta,0} = \left( c^\dag_{i,\up} c^\dag_{i+\delta,\dn} +
c^\dag_{i,\dn} c^\dag_{i+\delta,\up} \right)/\sqrt{2}$ are triplet
pair creation operators, and
$ \Delta^\dag_{i\delta} =  \left( c^\dag_{i,\up} c^\dag_{i+\delta,\dn} -
c^\dag_{i,\dn} c^\dag_{i+\delta,\up} \right) / \sqrt{2}$ 
is a singlet pair creation operator.
If one sets all the $W_i$ equal to $W$, the original $H_W$ interaction
 (\ref{eq:HW}) is recovered.
Here we will examine the effect of the individual terms.
$H_{W_1}$ renormalizes  the chemical potential and 
$H_{W_2}$ contains next-nearest and next-next-nearest neighbor one-electron 
hopping terms.
$H_{W_3}$ scatters an onsite singlet to neighbors sites while
$H_{W_4}$, which comes with a positive sign, is a triplet scattering term.
Finally $H_{W_5}$ and $H_{W_6}$ involve singlet pairs. 
It had been thought for the 2D system that the relevant terms for the quantum 
transition were $H_{W_5}$ and $H_{W_6}$. \cite{Assaad}

Here, in order to determine the effects of the individual terms, we have
studied the model on a two-leg ladder using DMRG techniques. 
All the runs were done on $2 \times 32$ ladders keeping up to 800 states leading 
to a maximum discarded weight of $10^{-6}$.
We calculated the singlet pairing correlation function
 $D_{\alpha\beta} (\ell)$ defined as
\begin{eqnarray}
 & & D_{xx} (\ell) = \langle \Delta_x(i+\ell) \Delta^\dag_x(i) \rangle \\
 & & D_{xy} (\ell) = \langle  \Delta_x(i+\ell) \Delta^\dag_y(i)   \rangle \\
 & & D_{yy} (\ell) = \langle   \Delta_y(i+\ell) \Delta^\dag_y(i)   \rangle 
\end{eqnarray}
where  $\Delta_\alpha(i)  =  c^\dag_{i,\up} c^\dag_{i+\delta_\alpha,\dn} -
c^\dag_{i,\dn} c^\dag_{i+\delta_\alpha,\up} $,
$\delta_x = (1,0)$ and $\delta_y = (0,1)$.
For clarity, in the following we show the rung-rung correlation function  
$D_{yy} (\ell)$. 
$D_{xx}(\ell)$ and $D_{yy}(\ell)$ were always positive while $D_{xy}(\ell)$ 
was negative corresponding to a $d_{x^2-y^2}$-like strucure.

\begin{figure}[htb]
 \begin{center}
  \epsfig{file=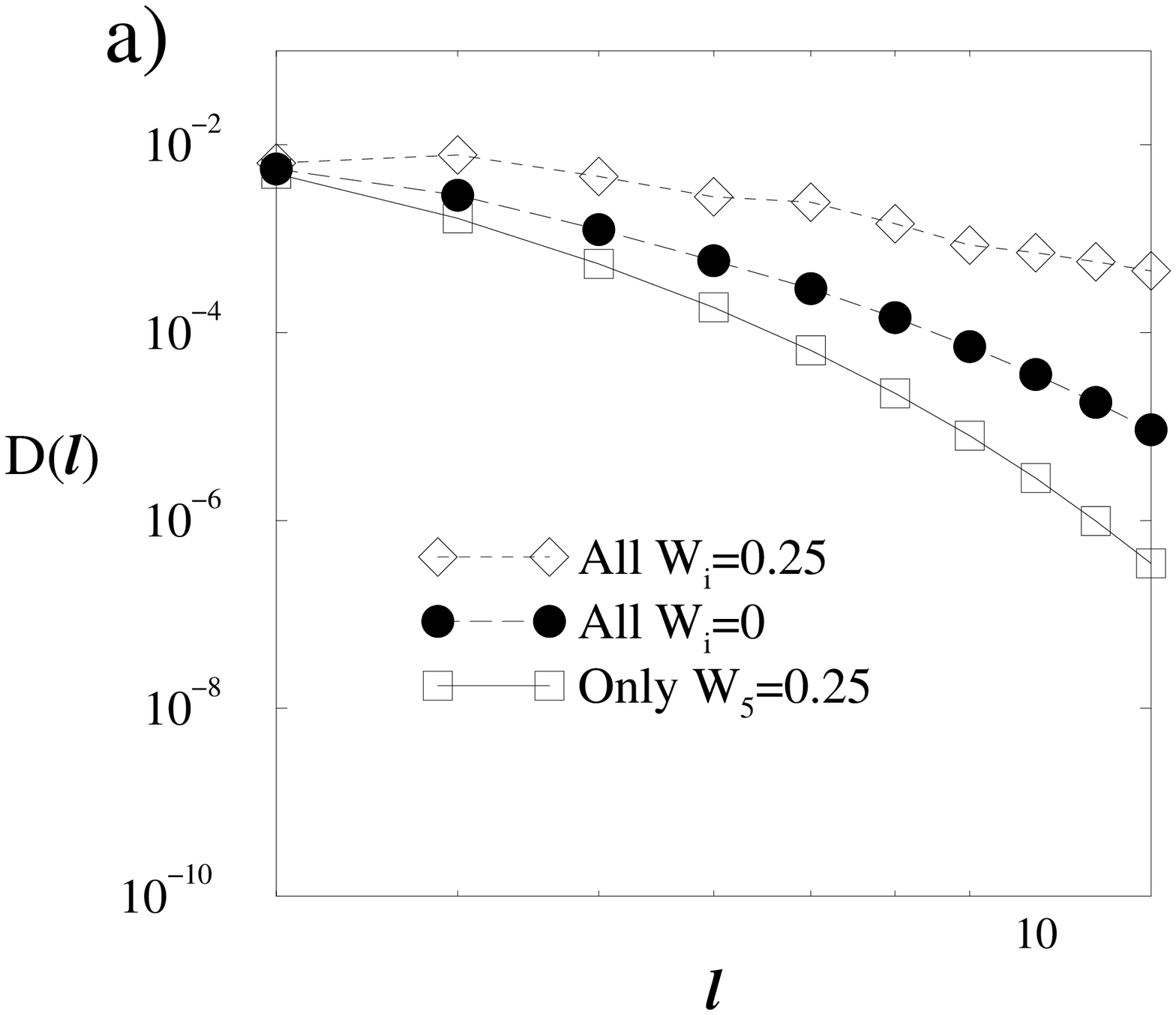,width=7cm}
  \epsfig{file=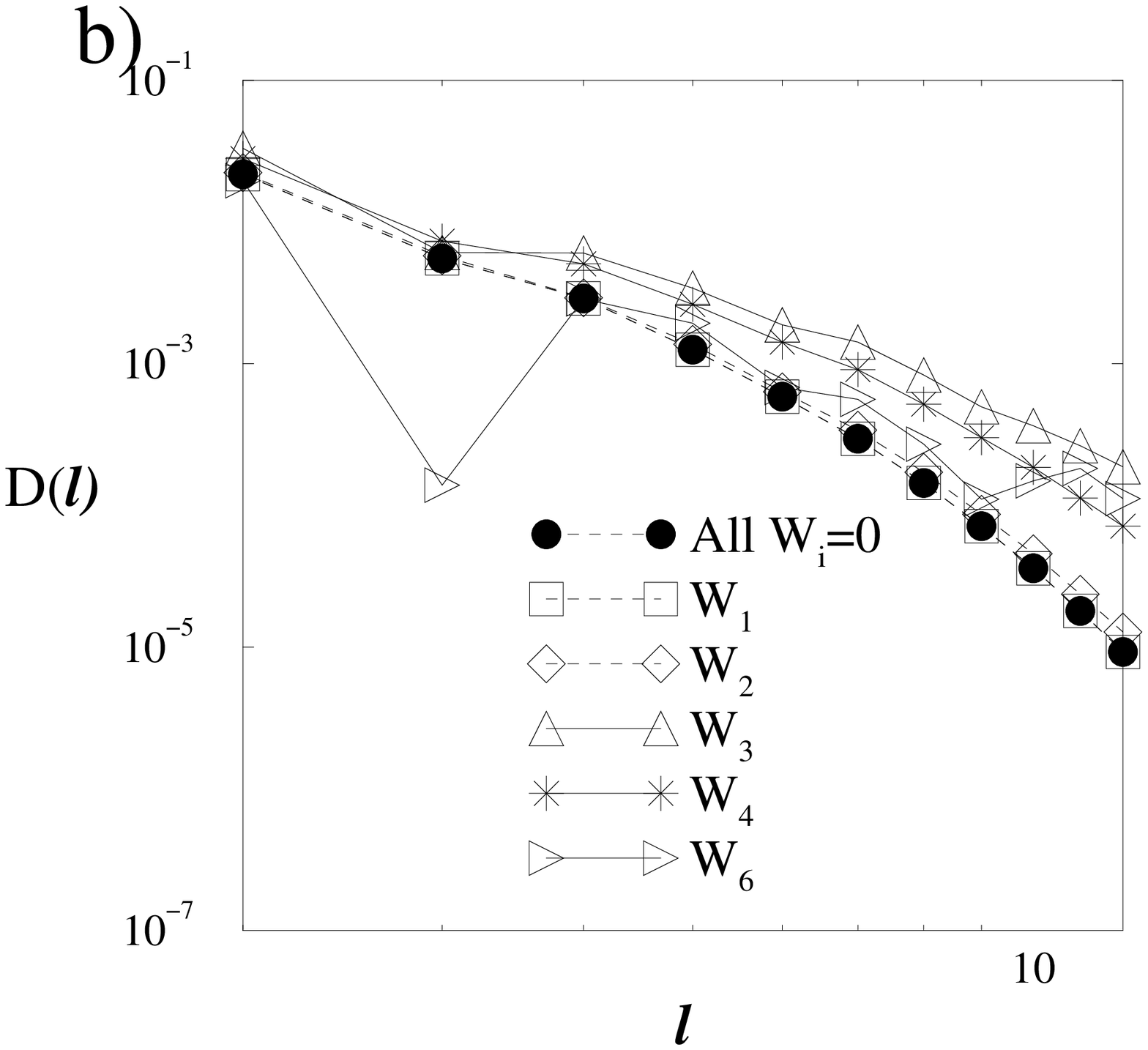,width=7cm}
 \end{center}
 \caption{The rung-rung singlet pairing correlation function $D_{yy}(\ell)$ 
versus $\ell$ on a half-filled ladder with $U=4$ for (a) all $W_i=0$, 
only $W_5=0.25$ and all $W_i=0.25$ and (b) all $W_i=0$ and only one $W_i=0.25$. }
\label{fig:half_filling_sc}
\end{figure}

The results for the half-filled case with $U=4$ and $W_i=0$ or 0.25
are shown in Fig. \ref{fig:half_filling_sc}. 
In the plot of $D_{yy}(\ell)$ we have kept $\ell \leq 12$, with the 
measurements made in the central portion of the ladder.
In this region the effects of the open ends are negligible.
We clearly see in part (a) that when all $W_i$ are turned on there is an
enhancement of the pairing (as found in the 2D Monte-Carlo simulations). 
However, if we only turn on $W_5$, there is a suppression of pairing.
For the 2-leg ladder, this can be understood by noting that $H_{W_5}$ can be 
written as an antiferromagntic exchange interaction
\begin{equation}
   H_{W_5} = 2W_5 \sum_{<ij>} \left( {\bf S}_i {\bf S}_{j} 
            -\frac{1}{4} n_i n_j   \right).
\label{eq:W5_exchange}
\end{equation}
Now as one knows, \cite{SpinGap} a 2-leg Heisenberg ladder has a spin gap
$\Delta_s \approx 0.51J$. 
Thus the effect of $H_{W_5}$ at half-filling is to increase the spin gap
by a factor of order $W_5$ and 
this leads to an exponentially
 more rapid decay of the pairing correlations.

Fig. \ref{fig:half_filling_sc} (b) shows the effect of the other
terms. 
$H_{W_1}$ has no effect as expected since it just renormalizes the chemical
potential and we have fixed $\langle n  \rangle=1$.
The additional one electron hopping term $H_{W_2}$ leads to only a small 
change in the pairing.
However $H_{W_3}$ which scatters an onsite singlet to neighbors sites 
enhances the pairing despite the presence of $U$ which lowers
the double occupancy.

\begin{figure}[htb]
 \begin{center}
  \epsfig{file=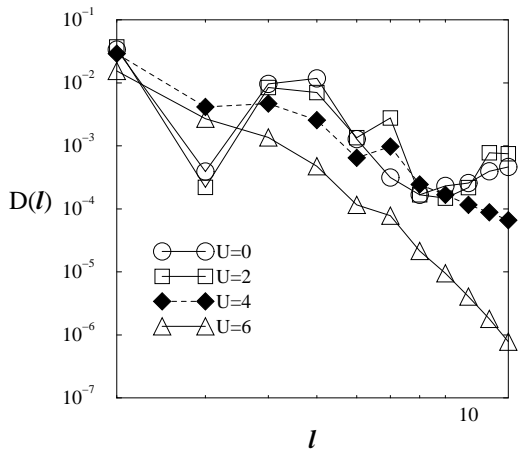,width=7cm}
 \end{center}
 \caption{ The rung-rung singlet pairing correlation function    
$D_{yy}(\ell)$ versus
    $\ell$ on a half-filled ladder with $W_3=0.25$ and various values of $U$. }
\label{fig:w3}
\end{figure}

\begin{figure}[htb]
 \begin{center}
  \epsfig{file=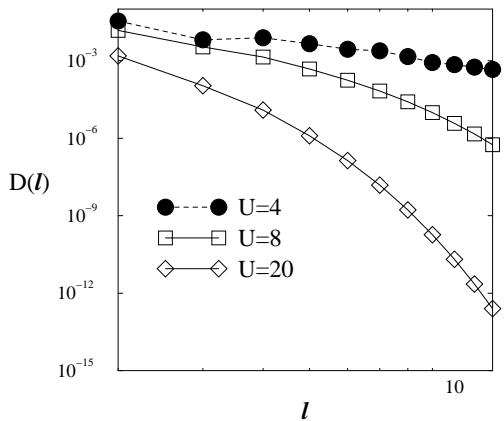,width=7cm}
 \end{center}
 \caption{The rung-rung singlet pairing correlation function 
$D_{yy}(\ell)$ versus
$\ell$ on a half-filled ladder with all $W_i=0.25$ and various values of $U$. }
\label{fig:Wall_U}

We now investigate the nature of the magnetic ordering 
by examining the spin--spin correlation function
\begin{equation}
 S(r) = \langle S_\ell^+S_{\ell+r}^-\rangle
\end{equation}
where $S_\ell^+$ ($S_\ell^-$) are the spin raising (lowering) operators
corresponding to ${\bf S}_\ell  = c^\dag_{\ell s} {\bf \sigma}_{ss'} 
c_{\ell s} $.
We then perform a Fourier transform to obtain the static
structure factor
\begin{equation}
   S(q) = \sum_r e^{iqr} S(r). 
\end{equation}
The resulting structure factor is plotted in Fig. \ref{fig:sq_n64_tot}
for various $W_i$ interactions.
As previously noted, the effect of $W_5$ is to increase the spin gap which
leads to a broadening of the Lorentzian and a decrease of its peak at
$(\pi,\pi)$.
When all of the $W_i$ terms are present, $S(q)$ appears to simply be reduced
in magnitude at all $q$ values indicating a reduction of the local moment
$\sqrt{\langle {\bf S }_\ell^2 \rangle }$ due to the delocalization effects
of $H_W$.
\end{figure}

We have performed other calculations including only $H_{W_3}$  
which show that for larger $U$ this enhancement is supressed 
as one would expect (see Fig. \ref{fig:w3}).
Nevertheless, for $U/t=4$ where the previous 2D Monte Carlo calculations were
run, $H_{W_3}$ contributes to enhanced the pairing.
$H_{W_4}$ also leads to enhanced singlet pairing.
Note that it has a positive coefficient which suppresses triplet pairing,
leaving more phase space for singlet pairing.
Finally $H_{W_6}$, which describes singlet pair hopping for 
$(i,\delta)$ to $(i,\delta')$, also enhances the pairing.
We should point out that although there is an enhancement in the pairing, 
this enhancement is in fact relatively small for the two-leg ladder
for reasonable values of $W$ and as $U$ is increased, so that the $W_3$
term is reduced, the suppression of the pairing by the $W_5$ term becomes
dominant. 
This is clearly seen in Fig. \ref{fig:Wall_U}, where we show 
the pairing correlation function for the half-filled ladder with all 
$W_i=0.25$ and various values of $U$.

\begin{figure}[htb]
 \begin{center}
  \epsfig{file=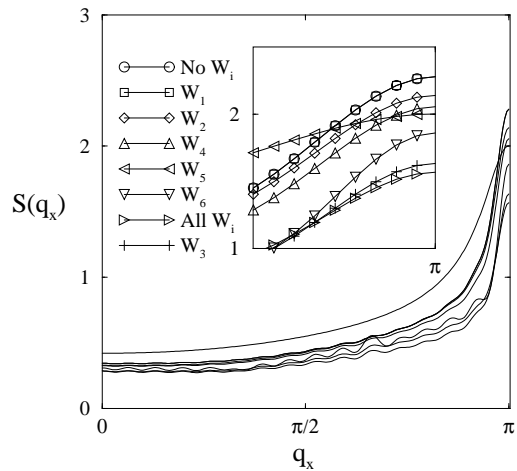,width=7cm}
 \end{center}
 \caption{Structure factor $S(q)$ of the spin--spin correlation
function versus $q_x$ for the half--filled system and $q_y=\pi$.
The inset shows an enlargement of the region near $(\pi,\pi)$.}
\label{fig:sq_n64_tot}
\end{figure}

\begin{figure}[htb]
 \begin{center}
  \epsfig{file=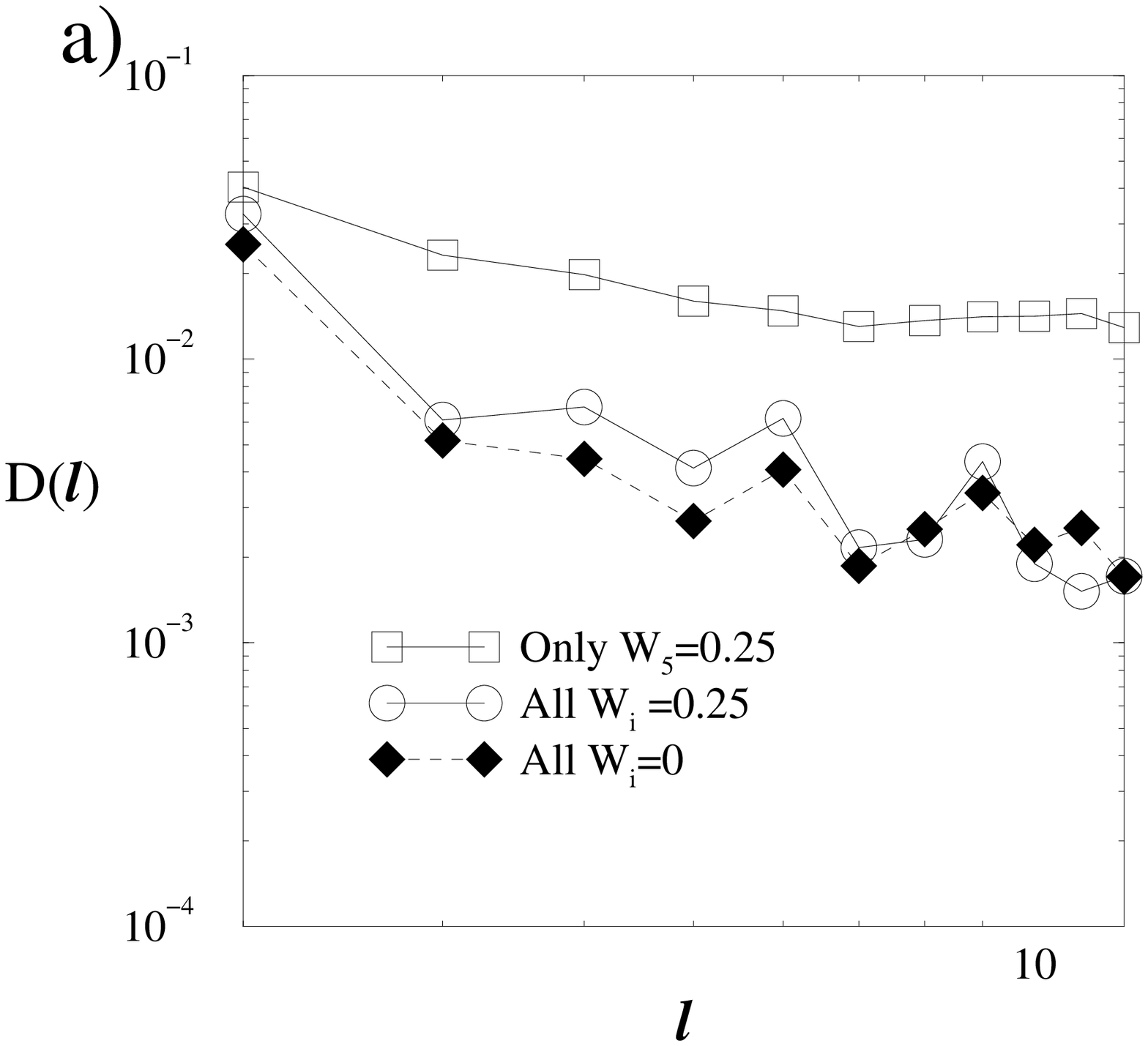,width=7cm}
  \epsfig{file=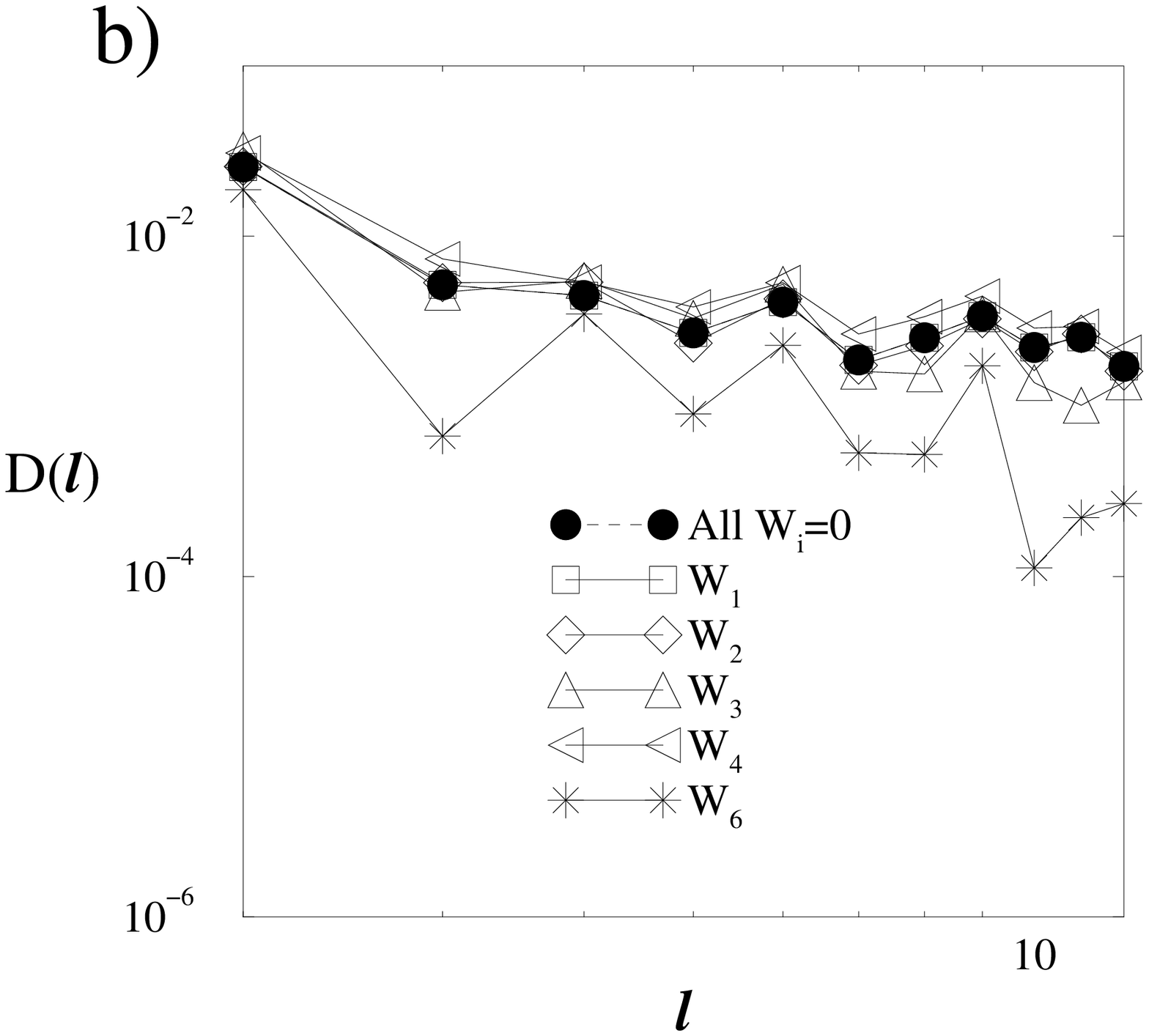,width=7cm}
 \end{center}
 \caption{The rung-rung singlet pairing correlation function 
$D_{yy}(\ell)$ versus
$\ell$ on a doped ladder with $\langle n \rangle =0.875$ and $U=4$  
for (a) all $W_i=0$, only $W_5=0.25$ and all $W_i=0.25$ and 
(b) all $W_i=0$ and only one $W_i=0.25$.}
\label{fig:doped_sc}
\end{figure}

We now turn to the doped case and consider the same lattice with $U=4$ and 
8 holes corresponding to a filling $\langle n \rangle = 0.875$.
Fig. \ref{fig:doped_sc} show results for $D_{yy}(\ell)$.
We clearly see in part (a) that in this case the inclusion of $H_{W_5}$ 
enhances the pairing while in part (b) we see that all of the remaining terms 
are essentially irrelevant.
Thus the $W_5$ term, which corresponds to adding a near neighbor exchange
$J = 2W_5$ enhances the pairing correlation in the doped system.

Thus we conclude, that while $H_W$ with $U=4t$ can slightly enhance
the pairing correlations of a half-filled ladder, this is in fact a small effect.
Furthermore, for large values of $U/t$, $H_W$ leads to a suppression
of the half-filled pairing correlations. 
This can be understood in terms of the dominance of $H_{W_5}$, which
represents an effective antiferromagnetic exchange increasing
the spin gap and suppressing the pairing correlations.
However, for the doped ladder, $W_5$ acts to enhance the pairing correlation 
since it increases the effective exchange interaction and the pair binding
energy.
Clearly, in light of the 2D results, where it was found that $H_W$ could
lead at half-filling to a $d_{x^2-y^2}$ pairing state, one would like to 
extend the DMRG calculations to a 3-leg ladder which has a vanishing spin gap
at half-filling.

We wish to thank D. Duffy and M. Fischer for helpful discussions. 
SD acknowledges support from the Swiss National Science Foundation. 
DJS and SRW wish to ackowledge the support from the US Department of
Energy under Grant No. DE-FG03-85ER45197.


\end{document}